\documentclass[12pt]{article}

\usepackage{amsfonts}

\begin{document}



\def\Bid{{\mathchoice {\rm {1\mskip-4.5mu l}} {\rm
{1\mskip-4.5mu l}} {\rm {1\mskip-3.8mu l}} {\rm {1\mskip-4.3mu l}}}}


\newcommand{\beq}{\begin{equation}}
\newcommand{\eeq}{\end{equation}}
\newcommand{\bea}{\begin{eqnarray}}
\newcommand{\eea}{\end{eqnarray}}
\newcommand{\eL}{{\cal L}}
\newcommand{\half}{\frac{1}{2}}
\newcommand{\J}{\bf J}
\newcommand{\bP}{\bf P}
\newcommand{\G}{\bf G}
\newcommand{\K}{\bf K}
\newcommand{\M}{{\cal M}}
\newcommand{\bu}{\bf u}
\newcommand{\la}{\lambda}
\newcommand{\ohalf}{\textstyle{1 \over 2}}
\newcommand{\thd}{\textstyle{1 \over 3}}
\newcommand{\tthd}{\textstyle{2 \over 3}}

\begin{flushright}
DOE-ER-40757-124\\
UTEXAS-HEP-99-2
\end{flushright}

\hfil\hfil

\begin{center}
{\Large {\bf Geometric Phases for Three State Systems}}
\end{center}
\begin{center}
\today
\end{center}
\hfil\break
\begin{center}
{\bf Mark Byrd \footnote{mbyrd@physics.utexas.edu}\\}
\hfil\break
{\it Center for Particle Physics \\
University of Texas at Austin \\
Austin, Texas 78712-1081}
\end{center}
\hfil\break


\begin{abstract}
The adiabatic geometric phases for general three state systems are discussed.  
An explicit parameterization for space of states of these systems is given.  
The abelian and non-abelian connection one-forms or vector potentials 
that would appear in a three dimensional quantum system with adiabatic 
characteristics are given explicitly.  This is done in terms of the 
Euler angle parameterization of $SU(3)$ which enables a straight-forward 
calculation of these quantities and its immediate generalization.
\end{abstract}



\pagebreak




\section{Introduction}

Geometric phases have received a great deal of attention since their 
description by Berry \cite{Berry}.  The reasons are clear.  They 
are a fundamental 
property of many quantum mechanical systems.  They also have a beautiful 
description in terms of differential geometry and fiber bundles
\cite{Simon} which 
is directly related to gauge field theory (see for example \cite{Nakahara}).  
Their physical importance 
was known long before the excitement about them in the mid 80's
\cite{Pan}, \cite{mandt}, \cite{LH}.  In 
spite of all the attention, there have been few worked out examples and the 
examples that have been worked out, the descriptions haven't been 
straight-forward.  The most well-known example is that of a two state 
system, namely a magnetic dipole, in a magnetic field.  This was 
the original example given by Berry \cite{Berry}.  Wilczek and Zee 
originally pointed out that there could exist non-
abelian geometric phases \cite{WandZ}.  Later people studied fermionic systems 
with a quadrapole Hamiltonian \cite{sadun}.  

Uhlmann later developed machinery, namely a parallel transport
\cite{Uhlmann}, for 
describing the non-abelian geometric phases associated with density 
matrices.  However, this was never applied to three state systems.
  Arvind {\it et al} and Khanna {\it et al} studied the 
geometric phases for three state 
systems that involve pure state density matrices \cite{m1}, \cite{m2} with a 
parameterization that was somewhat ad hoc.  
Mostafazadeh looked at a way of calculating the non-abelian 
geometric phases for a three state system with 
a two-fold degeneracy \cite{Ali} also with ad hoc coordinates.  These 
topics will be brought together here.  

In this paper the objective is to use explicit $SU(3)$ representations to 
extend and/or simplify 
several aspects of three state systems.
\begin{enumerate}
\item{The expression for the density matrix for three state systems.}
\item{The identification of the parameter spaces of these systems.}
\item{The calculation of the abelian geometric phases for three state 
systems.}
\item{The calculation of the non-abelian geometric phases of three 
state systems with a two-fold degeneracy.}
\end{enumerate}

An obvious example of a three state system would be a spin one particle 
in a magnetic field.  If this external magnetic field is ``slowly'' 
rotating then we may have the conditions for an adiabatic change in 
phase.  For a proper description of what ``slowly'' means in this 
context see \cite{Ali2}.  Unless otherwise stated, this paper will 
be concerned with the adiabatic geometric phases although with 
some work this could be extended to non-adiabatic phase changes.


\section{The Density Matrix for a System with Three Quantum States}

As is demonstrated in the next section the density matrix can 
be parameterized by the action of an $SU(3)$ transformation.  This 
will prove convenient for many calculations and is immediately 
generalizable to a system with an arbitrary number of states.  See 
\cite{us} and below for a discussion.  

The density matrix for general pure state three-level 
systems is given in \cite{m1} and \cite{m2}.  It can be represented in the 
following way: Let $\psi$ be a state in a three dimensional complex Hilbert 
space ${\cal H}^{(3)}$.  The density matrix is the matrix $\rho$ described 
by (in analogy with two state systems):
\begin{equation}
\rho = \psi \psi^{\dagger} = |\psi\rangle \langle \psi| = 
\frac{1}{3}(1+\sqrt{3}\vec{n}\cdot \vec{\lambda})
\label{algdensmatrix}
\end{equation}
$$
\psi \in {\cal H}^{(3)}, \;\;\;\;\;\;\;\;\;\;\;\;\;\; (\psi,\psi) = 1.
$$
Here the dagger denotes the hermitian conjugate, $\vec{n}$ is a real eight 
dimensional unit vector, $\vec{\lambda}$ represents the eight Gell-Mann 
matrices.
\begin{table}
$$
\begin{array}{crcr}

\lambda_1 = \left( \begin{array}{crcl}
                     0 & 1 & 0 \\
                     1 & 0 & 0 \\
                     0 & 0 & 0   \end{array} \right), &

\lambda_2 = \left( \begin{array}{crcr} 
                     0 & -i & 0 \\
                     i &  0 & 0 \\
                     0 &  0 & 0   \end{array} \right), &

\lambda_3 =  \left( \begin{array}{crcr} 
                     1 &  0 & 0 \\
                     0 & -1 & 0 \\
                     0 &  0 & 0   \end{array} \right), \\

\lambda_4 =  \left( \begin{array}{clcr} 
                     0 & 0 & 1 \\
                     0 & 0 & 0 \\
                     1 & 0 & 0   \end{array} \right), &

\lambda_5 = \left( \begin{array}{crcr} 
                     0 & 0 & -i \\
                     0 & 0 & 0 \\
                     i & 0 & 0   \end{array} \right), &

 \lambda_6 = \left( \begin{array}{crcr} 
                     0 & 0 & 0 \\
                     0 & 0 & 1 \\
                     0 & 1 & 0   \end{array} \right), \\

\lambda_7 = \left( \begin{array}{crcr} 
                     0 & 0 & 0 \\
                     0 & 0 & -i \\
                     0 & i & 0   \end{array} \right), &

\lambda_8 = \frac{1}{\sqrt{3}}\left( \begin{array}{crcr} 
                     1 & 0 & 0 \\
                     0 & 1 & 0 \\
                     0 & 0 & -2   \end{array} \right).

\end{array}
$$
\caption{The Gell-Mann Matrices}
\end{table}
 The dot product is the ordinary sum over repeated indices $n^r \lambda_r$.  
The $(\cdot,\cdot)$ is the inner product on the space ${\cal H}^{(3)}$.  
The pure state density matrix satisfies:
$$
\rho^{\dagger} = \rho^2 = \rho \geq 0 \;\;\;\;\; \mbox{Tr}\rho = 1.
$$
This is equivalent to the following conditions on $n$:
\begin{equation}
n^*=n \;\;\;\;\; n \cdot n =1 \;\;\;\;\; n \star n = n.
\label{ns}
\end{equation}
The star product is defined by
\begin{equation}
(a \star b)_i = \sqrt{3}d_{ijk}a_j b_k
\end{equation}
where the $d_{ijk}$ are the components of the completely symmetric tensor 
appearing in the anticommutation relations
$$
\{\lambda_i,\lambda_j\} = \frac{4}{3}\delta_{ij} + 2 d_{ijk} \lambda_k.
$$
Explicitly the nonzero $d_{ijk}$ are
$$
d_{118}= d_{228}= d_{338}= -d_{888}= \frac{1}{\sqrt{3}}\;\;\;\;\; d_{448}= d_{558}= d_{668}= d_{778}=-\frac{1}{2\sqrt{3}}
$$
$$
d_{146}= d_{157}= -d_{247}= d_{256}= d_{344}= d_{355}
= -d_{366}= -d_{377}=\half.
$$


\section{Parameter Spaces for Three State Systems}


The parameter space of states of the three state systems can 
easily be seen to be coset spaces of $SU(3)$.  The Euler angle 
parameters are a particularly convenient way in which to see this.

A representation of the coset space $SU(3)/U(2)$ and of the 
density matrix for the pure states of a three state system 
may be obtained in terms 
of the Euler parameters given in \cite{me}.  There the group $SU(3)$ 
is parameterized by
$$
D(\alpha,\beta,\gamma,\theta,a,b,c,\phi) = e^{(i\la_3 \alpha)} e^{(i\la_2 \beta)}
 e^{(i\la_3 \gamma)} e^{(i\la_5 \theta)} e^{(i\la_3 a)} e^{(i\la_2 b)} 
e^{(i\la_3 c)} e^{(i\la_8 \phi)}.
$$
With this parameterization and the explicit representation of the 
corresponding adjoint representation in terms of the Euler angle 
parameters in \cite{me2}, a parameterization of the density matrix 
of the three state system may be obtained by the following 
projection which is analogous to the Hopf map given in \cite{gandp}:
\begin{equation}
x = \pi(D) = D\left[\frac{1}{3}(1 - \sqrt{3}\lambda_8)\right]D^{-1}.
\label{proj}
\end{equation}
Here $x \in SU(3)/U(2)$ and $D$ represents a point in the space 
$SU(3)$.  This projection is clearly invariant under the right 
action of a $U(2)$ operation defined by $U \in U(2)$ with
$$
U = e^{(i\la_3 a^{\prime})} e^{(i\la_2 b^{\prime})} 
e^{(i\la_3 c^{\prime})} e^{(i\la_8 \phi^{\prime})}. 
$$
This then defines the projection from $SU(3)$ to $SU(3)/U(2)$.  Since 
the second term in equation (4) is simply an adjoint action on 
$\lambda_8$, it can be read directly from the equations given in 
\cite{me2}.  There the matrix $R_{ij}$ that satisfies 
\begin{equation}
D\lambda_iD^{-1} = R_{ij}\lambda_j
\label{adrep}
\end{equation}
 was given explicitly.

One may of course note that the projection operator is not unique.  
Any $3\times 3$ matrix with a one on its diagonal would be invariant 
under a $U(2)$ subgroup and would represent a pure state.  It is, 
however, rather convenient in this parameterization to use this 
particular matrix:
$$
\frac{1}{3}(1 - \sqrt{3}\lambda_8) = \left( \begin{array}{crcr} 
                     0 &  0 & 0 \\
                     0 &  0 & 0 \\
                     0 &  0 & 1   \end{array} \right),
$$
so that it is clear that the upper left $2 \times 2$ matrix of zeros 
will be unaffected by an $SU(2)$ transformation in that block.  This 
matrix could be substituted for another that has one $1$ on a diagonal 
and zeros elsewhere and still be invariant under (another) $SU(2)$.  
The invariance of this with respect to an overall phase gives the 
$U(2)$ invariance.  

Now equation (\ref{proj}) can be rewritten as
\begin{equation}
x = \left[\frac{1}{3}(1 - \sqrt{3}R_{8j}\lambda_j)\right] 
= \left[\frac{1}{3}(1+ \sqrt{3}n_j\lambda_j)\right]
\end{equation}
where we identify the $R_{8j}$ as the components of a vector that 
satisfies those properties given in equation (\ref{algdensmatrix}).  
This can be viewed 
as an arbitrary rotation of the vector $\la_8$ with an adjoint action 
of the group (equation (\ref{adrep})) and of course it is now clear 
that $x$ is identified with $\rho$.  

The vector $\vec{n}$ has the following components.

\begin{eqnarray}
n_1 &=& -R_{81} = -\frac{\sqrt{3}}{2} 
\cos 2\alpha \sin 2\beta \sin^2 \theta \nonumber \\
n_2 &=& -R_{82} =  \frac{\sqrt{3}}{2} 
\sin 2\alpha \sin 2\beta \sin^2 \theta \nonumber \\
n_3 &=& -R_{83} =  \frac{\sqrt{3}}{2} 
\cos 2\beta \sin^2 \theta \nonumber \\
n_4 &=& -R_{84} =  \frac{\sqrt{3}}{2} 
\cos(\alpha+\gamma) \cos \beta \sin 2\theta \nonumber \\
n_5 &=& -R_{85} = -\frac{\sqrt{3}}{2} 
\sin(\alpha+\gamma) \cos \beta \sin 2\theta \nonumber \\
n_6 &=& -R_{86} = -\frac{\sqrt{3}}{2} 
\cos(\alpha-\gamma) \sin \beta \sin 2\theta \nonumber \\
n_7 &=& -R_{87} = -\frac{\sqrt{3}}{2} 
\sin(\alpha-\gamma) \sin \beta \sin 2\theta \nonumber \\
n_8 &=& -R_{88} = -1+\frac{3}{2}\sin^2 \theta 
\end{eqnarray}

From this, using the equations $n_i = \psi^{\dagger}\lambda_i \psi$, it follows that
\begin{equation}
\psi = e^{i\chi} \left( \begin{array}{c}
                       e^{i(\alpha + \gamma)} \cos \beta \sin \theta \\
                       e^{-i(\alpha - \gamma)} \sin \beta \sin \theta \\
                       \cos \theta
                       \end{array} \right).
\end{equation}
This may be recognized as the third column of the $SU(3)$ matrix $D$ above, 
thus agreeing with the calculation given in \cite{m1} and coming full circle 
in the analysis.  In this case the overall phase $\chi$ may be 
identified as $-2\phi /\sqrt{3}$ in the matrix $D$.  In section 
\ref{nageophs} it will become clear why this works and it will 
be generalized for the case of non-abelian geometric phases.

Although many of the details have not been worked out for $SU(n)$ groups,
 (the Euler angle parameters, the adjoint representation, {\it etc}.) 
the method of identifying the space of the parameters is the same (see 
\cite{us}).  
For a system with $n$ states, one may express a general diagonal 
density matrix in terms of the squared elements of an $n-1$ sphere.  
Then to take it to a general basis, one acts with the appropriate $SU(n)$ 
matrix.  The result is always a subset of $SU(n)/T^{n-1}$, where 
$T^{n-1}$ is the maximal ($n-1$) torus for the group.  If there are 
degenerate eigenvalues in the matrix, this space is reduced.  For example in the 
case of three states discussed shortly, the parameter space is 
a subset of $SU(n)/(SU(2)\times U(1))$ since there exists a 
two-fold degeneracy.  For an $m$-fold degeneracy we reduce the space 
by $SU(m)$.  In the case of an adiabatic approximation, we will see 
this is a proper subset, but were we to relax this condition, the 
space would be isomorphic to these spaces, not subsets.  In this way, 
one may identify a necessary condition for non-abelian geometric 
phases, namely the existence of the degeneracy and thus an $SU(m)$ 
factor in the denominator of the above coset expression.

Using this parameterization one gains essentially 
nothing over the expression 
of the Bloch sphere for two-state systems.  In that case the common 
parameterization of the Bloch sphere,
$$
\left( \begin{array}{cc}
         a & 0 \\
         0 & 1-a
       \end{array} \right)
$$
is really no different than the one presented here,
$$
\left( \begin{array}{cc}
         \cos^2 \theta & 0 \\
         0 & \sin^2 \theta
       \end{array} \right)
$$
except that positivity is automatic.  However in the case of three state 
systems, we have
\begin{equation}
\left( \begin{array}{crcr}
         \cos^2 \theta \; \sin^2 \phi & 0 &  0 \\
         0 & \sin^2 \theta \; \sin^2 \phi & 0 \\
         0 & 0 & \cos^2 \phi 
       \end{array} \right).
\label{3densmat}
\end{equation}
This is a convenient parameterization since the analogous Bloch sphere 
would have parameters with a non-rectangular domain.  The parameterization 
given here (see also \cite{us}) then helps with the 
analysis of three state density matrices and their corresponding entropy 
\cite{meandmims}.


\section{Connection, Curvature, and Abelian Geometric Phases}

In the spirit and notation of Nakahara \cite{Nakahara}, we can now derive 
the connection one form, the curvature and the Geometric Phase of the 
three state system.  The connection one form, 
sometimes called Berry's connection, can be written in terms of $\psi$ 
in the following way.  Define the total phase to be
$$
\varphi \equiv i \int {\cal A}
$$
\begin{equation}
{\cal A} = {\cal A}_{\mu}dx^{\mu} = -i\langle \psi | d |\psi \rangle,
\end{equation}
where $d$ is the ordinary exterior derivative.  Using equation (7), this becomes
\begin{equation}
{\cal A} = d\chi + \sin^2 \theta [\cos^2 \beta (d\alpha +d\gamma) - \sin^2 \beta (d\alpha - d\gamma)]. 
\end{equation}
This agrees with reference \cite{m1} if the following identifications are made with those quantities on the left being those of reference \cite{m1} and those on the right being ours.
$$
\eta \leftrightarrow \chi \;\;\;\;\;\;\;\;
\theta \leftrightarrow \theta \;\;\;\;\;\;\; \chi{\lower5pt\hbox{\scriptsize 1}} \leftrightarrow \alpha + \gamma \;\;\;\;\;\;\; \chi{\lower5pt\hbox{\scriptsize 2}} \leftrightarrow \alpha - \gamma.
$$
The corresponding curvature two form is given by
\begin{eqnarray}
F &=& d{\cal A} = -i d\psi^{\dagger}\wedge d\psi \nonumber \\
  &=& \sin 2\theta \cos^2 \beta d\theta \wedge d(\alpha + \gamma)
      -\sin^2 \theta \sin 2 \beta d\beta \wedge d(\alpha + \gamma) \nonumber \\
  & & -\sin 2 \theta \sin^2 \beta d\theta \wedge d(\alpha - \gamma)
      -\sin^2 \theta \sin 2\beta d\beta \wedge d(\alpha - \gamma).
\end{eqnarray}
This then, is the analogue of the ``solid angle formula'' for the two state systems.  In other words, the integral of this curvature two form gives the geometric phase, just as 
$$
\varphi_g = \frac{1}{2}\Omega
$$
in two state systems, where $\Omega$ is the solid angle for the two sphere.  The geometric phase is just the integral of the connection one form without the overall phase factor $\chi$, that is,
\begin{eqnarray}
\varphi_g &=&\int \sin^2 \theta [\cos^2 \beta (d\alpha +d\gamma) - \sin^2 \beta (d\alpha - d\gamma)] \nonumber \\
          &=& \int [\sin^2 \theta \cos 2\beta d\alpha + \sin^2 \theta d\gamma],
\end{eqnarray}
which again, agrees with \cite{m1}.


\section{Non-abelian Geometric Phases}

\label{nageophs}

In this section a novel way of obtaining geometric phases for 3-state systems
 is given.  This method is a generalization and simplification over the 
method 
presented in \cite{Ali} and a generalization over the method given in the 
previous section.  The way the connection one-forms for the 3-state 
systems are derived here uses the fact that the state space of the system 
can be expressed in terms of the group $SU(3)$.  This enables the calculation 
of the forms without diagonalization of the Hamiltonian.  In effect, the 
Hamiltonian is taken to be in diagonal form initially.  It is then 
``undiagonalized'' by an $SU(3)$ action which takes it into a general 
non-diagonal hermitian matrix.  This method has the advantage of being 
potentially generalizable to other states, not just eigenstates of the 
Hamiltonian.  (Of course, one has to be careful of what the adiabatic 
assumption means then.  This is well described in \cite{Ali2}.)  
It also has the advantage of being generalizable to $SU(n)$.  
Whereas one does not have a way of finding the eigenvalues of an $n\times n$ 
matrix, one would be able to use $SU(n)$ matrices and derive the connection 
forms for an $n$-state system.  (Again, see \cite{us}.)

The aim is to find the adiabatic non-abelian geometric phase associated 
to the two-fold degeneracy of energy eigenvalues of the general
 Hamiltonian for a 3-state system.  These are the simplest 
non-abelian geometric phases.  

Let $H(t) = H(\vec{R}(t))$ be the time dependent Hamiltonian 
of the system and let $E_n(t)$ be its eigenvalues.  Then if the Hamiltonian 
is periodic in time with period $T$, {\it i.e.}, the curve 
$C$:$[0,T]\rightarrow M$ is closed.  Here $M$ is the manifold parameterized 
by the coordinates $\vec{R}$.  For the adiabatic approximation, $n$ 
labels the eigenstates, $|\psi\rangle$, of the Hamiltonian and does not 
change.  This means there is a unitary matrix $U(n)$ relating $|\psi(T)\rangle$ and 
$|\psi(0)\rangle$ which is given by
$$
e^{-\frac{i}{\hbar}\int_0^T E_n(t)}{\cal P}\left[e^{i\oint_C A_n}  \right].
$$
Here ${\cal P}$ is the path-ordering operator and $A_n$ is a Lie algebra 
valued (connection) one-form whose matrix elements are locally given by:
\begin{equation}
A_n^{ab} = i\langle n,a,\vec{R}|d|n,b,\vec{R}\rangle.
\label{nageoph}
\end{equation}

It is important to note that the Hamiltonian is a $3 \times 3$ Hermitian 
matrix which can be viewed as an element of the algebra of $SU(3)$, 
{\it i.e.},
$$
H(\vec{R}) = b \sum_{i=0}^{8}R^i\lambda_i,
$$
where $R^i$ are real parameters, the $\lambda_i$ are 
$\lambda_0 = \Bid_{3\times 3}$ and the Gell-Mann matrices of Table 
(1).  Here the constant $b$ is taken to be one.  
The adiabaticity assumption may then be expressed as $T>\!\!>1$.  

The Hamiltonian, $H$, can be expressed in terms of the diagonalized 
Hamiltonian, $H_D$.
$$
H(\vec{R}) = U(\vec{R})H_DU^{-1}(\vec{R}),
$$
where $U(\vec{R}) \in SU(3)$ and 
$$
H_D = \left( \begin{array}{crcr}
                                E_1 &  0  &  0  \\
                                0  & E_1 &  0  \\
                                0  &  0  & E_3 
              \end{array} \right).
$$
In this form it is obvious that $M \subset \mathbb{C}$P$^2$ and what is more, 
it is clear from (\cite{me}) that only the angles $\alpha, \beta, \gamma$ and 
$\theta$ will remain since $\lambda_1, \lambda_2, \lambda_3$ and $\lambda_8$ 
commute with $H_D$.  Explicitly, the Hamiltonian in undiagonalized form, 
$H$, is given by
\begin{eqnarray}
H_{11} &=& E_1(\cos^2 \beta\cos^2\theta+\sin^2\beta)
              +  E_3\cos^2\beta\sin^2\theta       \nonumber   \\
H_{12} &=& (E_1-E_3)e^{-2i\alpha}\cos\beta\sin\beta\sin^2\theta   \nonumber  \\
H_{13} &=& (E_3-E_1)e^{-i(\alpha+\gamma)}\cos\beta\sin\theta\cos\theta 
\nonumber\\
H_{21} &=& (E_1-E_3)e^{2i\alpha}\cos\beta\sin\beta\sin^2\theta   
\nonumber     \\
H_{22} &=& E_1(\sin^2 \beta\cos^2\theta+\cos^2\beta)
              +  E_3\sin^2\beta\sin^2\theta    
\nonumber  \\
H_{23} &=& (E_1-E_3)e^{i(\alpha-\gamma)}\sin\beta\sin\theta\cos\theta 
\nonumber\\
H_{31} &=& (E_3-E_1)e^{i(\alpha+\gamma)}\cos\beta\sin\theta\cos\theta 
\nonumber \\
H_{32} &=& (E_1-E_3)e^{-i(\alpha-\gamma)}\sin\beta\sin\theta\cos\theta 
\nonumber\\
H_{33} &=&  E_1\cos^2\theta +E_3\sin^2\theta  \nonumber
\end{eqnarray}
It can easily be shown that these angles parameterize $\mathbb{C}$P$^2$.  
In this way one can easily identify the patches needed for certain 
circumstances.  This is analogous to the calculation here.

As is well known, the matrix that diagonalizes $H$ is composed of its 
eigenvectors.  Therefore, given that $H=UH_DU^{-1}$, $H_D=U^{-1}HU$, 
so we have our $|\psi\rangle$s, the eigenvectors of $H$, they are
$$
\left( \begin{array}{c}
e^{-i(\alpha + \gamma)}\cos\beta \cos\theta \\
-e^{i(\alpha - \gamma)}\sin\beta \cos\theta \\
-\sin\theta
\end{array}\right),
\left( \begin{array}{c}
e^{-i(\alpha - \gamma)}\sin\beta  \\
e^{i(\alpha + \gamma)}\cos\beta   \\
0
\end{array}\right),
\left( \begin{array}{c}
e^{-i(\alpha + \gamma)}\cos\beta \sin\theta \\
-e^{i(\alpha - \gamma)}\sin\beta \sin\theta \\
\cos\theta
\end{array}\right).
$$
One can check that these are already orthonormal due to the fact that 
$U\in SU(3)$.

Now all that needs to be done is calculate the connection forms 
given by (\ref{nageoph}).  These are given by
$$
A_1 = \cos 2\beta \cos^2\theta \;d\alpha + \cos^2\theta \;d\gamma,
$$
and
$$
A_2 = \left( \begin{array}{crcr}
                      -\cos2\beta \;d\alpha - d\gamma & 
e^{-2i\gamma}[\sin 2\beta \sin \theta \;d\alpha - i \sin \theta \;d\beta] \\
e^{2i\gamma}[\sin 2\beta \sin \theta \;d\alpha + i \sin \theta \;d\beta] &
\cos 2\beta \sin^2 \theta \;d\alpha + \sin^2 \theta \;d\gamma
              \end{array} \right).
$$

This is a expression in terms of $SU(3)$ Euler angle coordinates.  We 
can generalize this by using the expression (\ref{3densmat}).  This allows 
us to express the density matrix for an $n$-state system in terms of the 
Euler angle coordinates and the components of the $n-1$ sphere along the 
diagonal and an overall scale factor.  Thus the eigenvalues need not 
be those of the Hamiltonian but of any observable.  Then a similar 
analysis holds for states that are not eigenvectors of the Hamiltonian 
but eigenvectors of another observable with the caution that, as stated 
before, one must be careful of what one means by an adiabatic approximation.


\section{Conclusions/Comments}

The diagonalized density matrices can be parameterized 
by the squared elements of 
the sphere (an $n-1$ sphere for a system of $n$ states) combined with 
an $SU(n)$ action.  This novel 
parameterization helps to identify the parameter spaces of these systems.  
The spaces are isomorphic to subspaces of $SU(n)/T^n$ for all eigenvalues 
unique, to subspaces of $SU(n)/(SU(2)\times T^{n-1})$ for one two-fold 
degeneracy, $SU(n)/(SU(3)\times T^{n-2})$ for one three-fold degeneracy, 
{\it etc.}  This is because the density matrix and Hamitonian are 
both in the algebra of the group and can represented as $UAU^\dagger$,
where $U\in SU(n)$ and $A$ is the non-diagonal density matrix or 
Hamiltonian.  When this is the Hamiltonian, we immediately know the 
eigenvectors because they are the rows of the matrix that diagonalizes 
the Hamiltonian.  This enables the evaluation (in principle) 
of the geometric phases 
for the $n$-state systems.  Here we have shown this 
explicitly for the case of three quantum states.  

In this analysis the 
Euler angle parameterization has been extremely useful and although 
its generalization to $SU(n)$ is possible, the decomposition into 
components of the spheres and $SU(n)$ actions is independent of the 
parameterization.

In \cite{Ali} applications to multi-pole Hamiltonians were discussed.  I would 
like to add that there are phenomenological nuclear physics 
models that use $SU(3)$.  These multi-pole Hamiltonians are expressible 
in terms of the differential operators in \cite{me}, and \cite{me2}.  The 
author expects to perform a further analysis of the relations to those and 
other multi-pole Hamiltonians in the near future.


\section*{Acknowledgements}

I would like to thank Alonso Botero, Luis J. Boya, Richard Corrado 
Mark Mims, and E. C. G. Sudarshan for many insightful discussions.  
I would like to thank DOE for its support under the grant number 
DOE-ER-40757-123.


\end{document}